\documentclass{PoS}

\newcommand{\clqcd}{CL\kern-.25em\textsuperscript{2}QCD}
\newcommand{\codename}{\clqcd}

\title{\codename\ - Lattice QCD based on OpenCL}

\ShortTitle{\codename\ - Lattice QCD based on OpenCL}

\author{Owe Philipsen, \speaker{Christopher Pinke$^\ast$,  Alessandro Sciarra}\\
        Institut für Theoretische Physik - Johann Wolfgang Goethe-Universität\\
        Max-von-Laue-Str. 1, 60438 Frankfurt am Main\\
        E-mail: \email{philipsen, pinke, sciarra @th.physik.uni-frankfurt.de} }

\author{{Matthias Bach}\\
        Frankfurt Institute for Advanced Studies / Institut für Informatik - Johann Wolfgang Goethe-Universität\\
				Ruth-Moufang-Str. 1, 60438 Frankfurt am Main \\
        E-mail: \email{bach@compeng.uni-frankfurt.de}}

\abstract{
We present the Lattice QCD application {\clqcd}, which is based on OpenCL and can be utilized to run on Graphic Processing Units as well as on common CPUs.
We focus on implementation details as well as performance results of selected features.
\clqcd\ has been successfully applied in LQCD studies at finite temperature and density and is available at \clqcdAdress.
}

\FullConference{The 32nd International Symposium on Lattice Field Theory\\
		 23-28 June, 2014\\
		 Columbia University New York, NY}

\usepackage{etex}

\usepackage{epsfig}
\usepackage{multicol}
\usepackage{pstricks,pst-grad}

\usepackage[english]{babel}
\usepackage[utf8]{inputenc}
\usepackage[T1]{fontenc}
\usepackage{amsmath, amssymb}
\usepackage{nicefrac}
\usepackage{bm}

\usepackage{graphicx}
\usepackage[format=hang, font=small,labelfont=bf,textfont=it, justification=centerlast, margin=1mm]{caption}
\usepackage{wrapfig}

\usepackage{booktabs}
\usepackage{multirow}

\usepackage{tikz}
\usetikzlibrary{shadows}
\usepackage{fancybox}
\usetikzlibrary{calc}
\usetikzlibrary{shapes}
\usepackage{pgfplots}
\pgfplotsset{every axis/.style={width=\textwidth}} 

\usepackage{scalefnt}

\usepackage{marginnote}

\usepackage{todonotes}

\usepackage{csquotes}

\PassOptionsToPackage{
	natbib=true,
        style=numeric-comp, 
	bibstyle=numeric-comp, 
        hyperref=true,
        backend=bibtex, 
        maxbibnames=2,
        firstinits=true,
        uniquename=init,
        maxcitenames=2,
        parentracker=true,
	uniquelist=false,
        url=false,
        doi=false,
        isbn=false,
        eprint=true,
        sorting=none
}   {biblatex}
\usepackage{biblatex}

\AtEveryCite{%
  }

\newcommand{\mminus}{\ensuremath{D^{-1}}}
\newcommand{\tmlqcd}{\texttt{tmLQCD}}

\newcommand{\loewe}{LOEWE	-CSC}
\newcommand{\sanam}{SANAM}

\newcommand{\dslash}{\ensuremath{{\not}D}}

\newcommand{\clqcdAdress}{{\textcolor{cyan!30!blue}{\texttt{http://code.compeng.uni-frankfurt.de/projects/clhmc}}}}

\newcommand{\myurl}[1]{\textcolor{black!60!blue}{\texttt{\scriptsize{#1}}}}
\newcommand{\cmake}{\textsc{CMake}}
\newcommand{\cppnamespace}[1]{\textcolor{black!40!cyan}{\texttt{#1}}}
\newcommand{\cppclass}[2]{\cppnamespace{#1::}\textcolor{black!60!green}{\texttt{#2}}}

\usepackage{footmisc}

\begin{document}
\author{\textcolor{white}{\speaker{Christopher Pinke, Alessandro Sciarra}}}
\vspace*{-1cm}

\section{Lattice QCD at Finite Temperature}

Lattice QCD (LQCD) successfully describes many aspects of the strong interactions and is the only method available to study QCD from first principles.
The idea is to discretize space-time on a $N_\sigma^{3}\times N_\tau$ hypercube  with lattice spacing $a$ and treat this system with numerical methods.
State-of-the-art lattice simulations require high-performance computing and constitute one of the most compute intensive problems in science.
The discretization procedure is not unique and several different lattice theories of QCD have been developed.
It is important, in general, to cross check each result using different formulations.

The QCD phase diagram is of great interest both theoretically and experimentally, e.g. at the dedicated programs at RHIC at Brookhaven, LHC at CERN or at the future FAIR facility in Darmstadt\footnote{See \myurl{\scriptsize{http://www.bnl.gov/rhic/}}, \myurl{\scriptsize{http://home.web.cern.ch/}}, and \myurl{\scriptsize{http://www.fair-center.de}}\ .}.
On the lattice, studies at finite temperature $T$ are possible via the identification $T = (a(\beta)N_\tau)^{-1}$.
Thus, scans in $T$ require simulations at multiple values of the lattice coupling $\beta$. 
In addition, to employ a scaling analysis, simulations on various spatial volumes $N_\sigma^{3}$ are needed (to avoid finite size effects one typically uses $N_\sigma/N_\tau \approx 3$).
Hence, studies at finite $T$ naturally constitute a parallel simulation setup.
Currently, these investigations are restricted to zero chemical potential $\mu$, as the {sign-problem} prevents direct simulations at $\mu>0$.
To circumvent this issue one can use reweighting, a Taylor series approach or one can employ a purely {imaginary chemical potential $\mu_{I}$}.

On the lattice, observables are evaluated by means of importance sampling methods by generating ensembles of gauge configurations $\{U_m\}$ using as probability measure the Boltzmann-weight $p[U, \phi] =\exp\left\{ - S_\text{eff}[U, \phi] \right\}$.
Expectation values are then
\begin{align*}
   \left< K \right>  \approx \frac{1}{N}\sum_m K[U_m]\;.
\end{align*}
These ensembles are commonly generated using the {Hybrid-Monte-Carlo (HMC)} algorithm \parencite{Duane:1987de}, which does not depend on any particular lattice formulation of QCD.

The fermions enter in the effective action $S_\text{eff}$ via the fermion determinant $\det D$, which is evaluated using pseudo-fermions $\phi$, requiring the inverse of the fermion matrix, \mminus.
The fermion matrix $D$ is specific to the chosen discretization.
The most expensive ingredient to current LQCD simulations is the inversion of the fermion-matrix
\begin{align*}
  D \phi = \psi \quad\Rightarrow\quad \phi = \mminus\; \psi \;,
\end{align*}
which is carried out with Krylov subspace methods, e.g. conjugate gradient (CG).
During the inversion, the matrix-vector product $D \phi$ has to be carried out multiple times.
The performance of this operation, like almost all LQCD operations, is {limited
by the memory bandwidth}.
For example, in the Wilson formulation, the derivative part of $D$, the so-called \dslash, requires to read and write 2880 Bytes per lattice site in each call, while it performs \emph{only} 1632 FLOPs per site, giving a rather low numerical density $\rho$ (FLOPs per Byte) of $\sim 0.57$. 
In the standard staggered formulation, the situation is even more bandwidth-dominated.
To apply the discretization of the Dirac operator on a fermionic field ($D_{KS}\;\phi$) 570 FLOPs per each lattice site are performed and 1584 Bytes are read or written, with a consequent smaller $\rho$ of $\sim 0.35$.
This emphasizes that LQCD requires hardware with a high memory-bandwidth to run effectively, and that a meaningful measure for the efficiency is the achieved bandwidth.
In addition, LQCD functions are local, i.e. they depend on a number of nearest neighbours only.
Hence, they are very well suited for parallelization.

\section{OpenCL and Graphic Cards}

{   \centering
   \scalebox{0.8}{\begin{tabular}{@{} l|c|c|c|c @{}}
        \toprule
        & \multicolumn{1}{c|}{\multirow{2}*{\textsc{Chip}}}        & \textsc{Peak SP}  & \textsc{Peak DP}  & \textsc{Peak BW}   \\[-0.5ex]
        &                 & \small\texttt{\{GFLOPS\}} & \small\texttt{\{GFLOPS\}} & \small\texttt{\{GB/s\}} \\
        \hline
        AMD Radeon HD 5870     & Cypress         & 2720     & 544      & 154       \\
        AMD Radeon HD 7970     & Tahiti          & 3789     & 947      & 264       \\
        AMD FirePro S10000      & Tahiti          &  2$\times$3410    & 2$\times$850    &   2$\times$240 \\
        \hline
        NVIDIA GeForce GTX 680 & Kepler          & 3090     & 258      & 192       \\
        NVIDIA Tesla K40 & Kepler          & 4290    &  1430     &   288     \\
        \hline
        AMD Opteron 6172       & Magny-Cours     & 202      & 101      & 43      \\
        Intel Xeon E5-2690     & Sandy Bridge EP & 371      & 186      & 51      \\
        \bottomrule
     \end{tabular}}
  \captionsetup{width=0.8\linewidth}
  \captionof{table}{Theoretical peak performance of current GPUs and CPUs. SP and DP denote single and double precision, respectively. BW denotes bandwidth. }
  \label{tab:cpu-gpu-theo-peak}
}

\vspace{5mm}

{ \centering
  \begingroup
  \renewcommand{\arraystretch}{1.09} 
  \scalebox{0.8}{\begin{tabular}{c|c|c|c}
	\toprule
	& \multicolumn{2}{c|}{\loewe} & \sanam \\
	\hline
	GPU nodes & 600 & 40 & 304\\
	GPUs/node & 1 $\times$ AMD 5870 & 2 $\times$ AMD S10000 & 2 $\times$ AMD S10000\\
	CPUs/node & 2 $\times$ Opteron 6172 & 2 $\times$ Intel Xeon E5-2630 v2 & 2 $\times$ Xeon E5-2650\\ 
	\bottomrule
  \end{tabular}}
  \endgroup
  \captionof{table}{AMD based clusters where \clqcd\ was used for production runs.}\label{tab:clusters}
}

\vspace{5mm}

Graphics Processing Units (GPUs) surpass CPUs in peak performance as well as in memory bandwidth (see \tablename~\ref{tab:cpu-gpu-theo-peak}) and can be used for general purposes. 
Hence, many clusters are today accelerated by GPUs, for example \mbox{\loewe} in Frankfurt \parencite{Bach2011a} or \sanam\ \parencite{bib:sanampaper} (see \tablename~\ref{tab:clusters}).
GPUs constitute an inherently parallel architecture.
As LQCD applications are always memory-bandwidth limited (see above) they can benefit from GPUs tremendously.
Accordingly, in recent years the usage of GPUs in LQCD simulations has increased.
These efforts mainly rely on CUDA as computing language, applicable to NVIDIA hardware \emph{only}\footnote{See  \myurl{\scriptsize{https://developer.nvidia.com/cuda-zone}}\hspace{1mm} and\hspace{1mm} \myurl{\scriptsize{https://github.com/lattice/quda}}\ for the {\bfseries{QUDA}} library.}.
A hardware independent approach to GPU applications is given by the  \textit{Open Computing Language} ({\bfseries OpenCL})\footnote{See \myurl{https://www.khronos.org/opencl}\ .}, which is an open standard to perform calculations on heterogeneous computing platforms. 
This means that GPUs and CPUs can be used together within the same framework, taking advantage of their synergy and resulting in a high portability of the software. 
First attempts to do this in LQCD have been reported in \parencite{Philipsen:2011sa}.

\begin{wrapfigure}{r}{0.33\hsize}
  \centering
  \includegraphics[width=0.99\hsize]{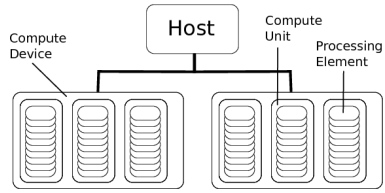}
    \captionof{figure}{OpenCL concept}\label{fig:ocl-concept}
\end{wrapfigure}

An OpenCL application consists of a \textit{host} program coordinating the execution of the actual functions, called \textit{kernels}, on \textit{computing devices} (\figurename~\ref{fig:ocl-concept}), like for instance GPUs or a CPUs.

Although the hardware has different characteristics, GPU programming shares many similarities with parallel programming of CPUs.
A computing device consists of multiple \emph{compute units}.
When a kernel is executed on a computing device, actually a huge number of kernel instances is launched.
They are mapped onto \emph{work-groups} consisting of \emph{work-items}.
The work-items are guaranteed to be executed concurrently only on the processing elements of the compute unit (and share processor resources on the device).

Compared to the main memory of traditional computing systems, on-board memory capacities of GPUs are low, though increasing more and more\footnote{For instance, the GPUs given in \tablename~\ref{tab:clusters} have on-board memory capacities of 1 GB and $2\times6$ GB, respectively, on \loewe\ and $2\times3$ GB on \sanam.}.
This constitutes a clear boundary for simulation setups.
Also, communication between host system and GPU is slow, limiting workarounds in case the available GPU memory is exceeded.
Nevertheless, as finite $T$ studies are usually carried out on moderate lattice sizes (in particular $N_\sigma \gg N_\tau$), this is less problematic for the use cases \clqcd\ was developed for.

\section{\clqcd\ Features}

\clqcd\ is a Lattice QCD application based on OpenCL, applicable to CPUs and GPUs.
Focusing on Wilson fermions, it constitutes the first such application for this discretization type \parencite{Bach:2012iw}.
In particular, the so-called { Twisted Mass Wilson fermions} \parencite{Shindler:2007vp, Frezzotti:2003ni}, which ensure $\mathcal{O}(a)$ improvement at maximal twist, are implemented.
Recently, the (standard) formulation of {{staggered fermions}} has been added.
Improved gauge actions and standard inversion and integration algorithms are available, as well as ILDG-compatible IO\footnote{Via LIME, see \myurl{http://usqcd.jlab.org/usqcd-docs/c-lime}\ .} and the RANLUX Pseudo-Random Number Generator (PRNG) \parencite{Luescher1994100}.
More precisely, \clqcd\ provides the following executables.
\begin{itemize}
\item {\bfseries{HMC:}} Generation of gauge field configurations for $N_f=2$ Twisted Mass Wilson type or pure Wilson type fermions using the HMC algorithm \parencite{Duane:1987de}.
\item {\bfseries{RHMC:}} Generation of gauge field configurations for $N_f$ staggered type fermions using the Rational HMC algorithm \parencite{ClarkKennedy:RHMC}.
\item {\bfseries{SU3HEATBATH:}} Generation of gauge field configurations for $SU(3)$ Pure Gauge Theory using the heatbath algorithm \parencite{Creutz:1980zw, Cabibbo:1982, Kennedy:1985}.
\item {\bfseries{INVERTER:}} Measurements of fermionic observables on given gauge field configurations.
\item {\bfseries{GAUGEOBSERVABLES:}} Measurements of gauge observables on given gauge field configurations.
\end{itemize}

\section{\clqcd\ Code Structure}

The host program of \clqcd\ is set up in \texttt{C++}, which allows for independent program parts using \texttt{C++} functionalities and also naturally provides extension capabilities. 
Cross-platform compilation is provided using the \cmake{} framework.\footnote{See \myurl{http://www.cmake.org}\ .}

\begin{figure}[h]
 \centering
 \includegraphics[width=\hsize]{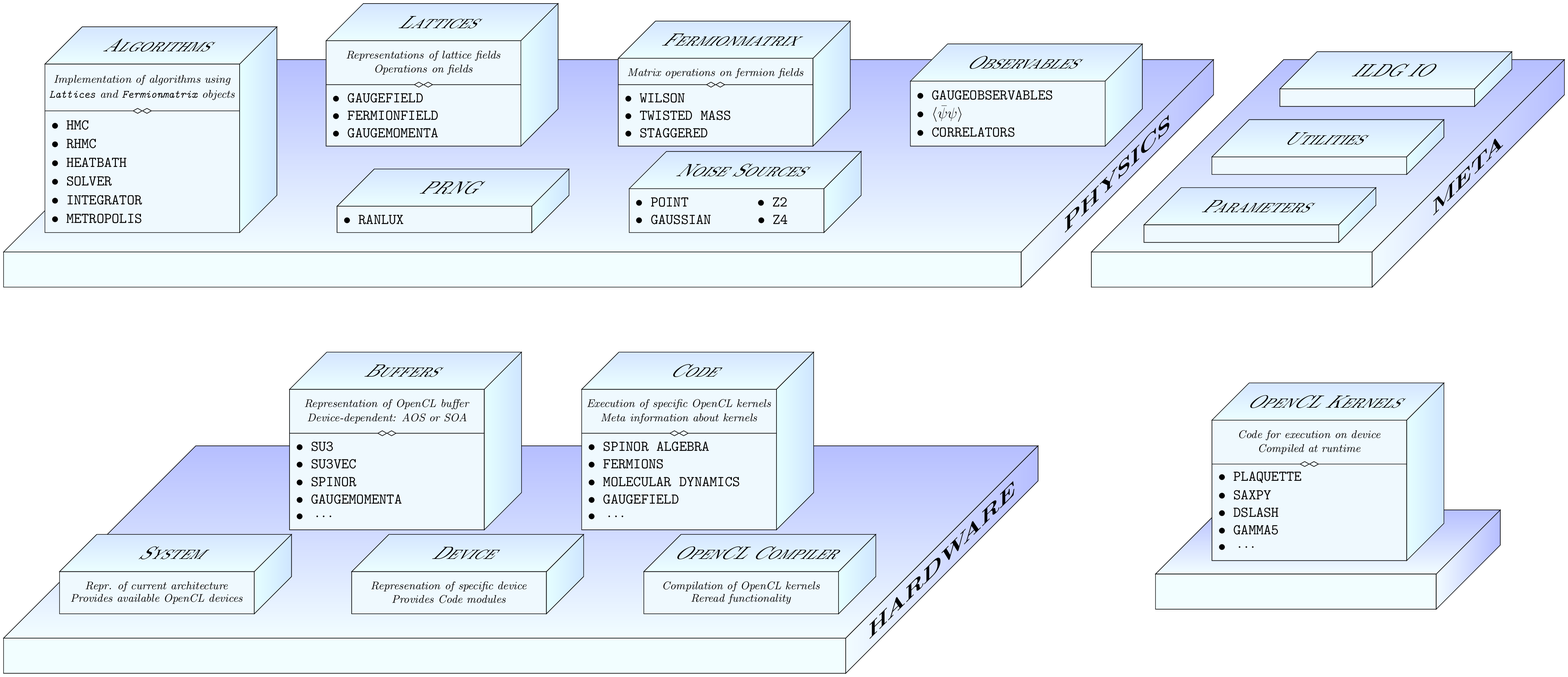}
 \captionof{figure}{\clqcd\ code structure. Packages and substructures are realized as namespaces. The names of the various components are, for the sake of simplicity, not always identical to those in the code.}\label{fig:code-structure}
\end{figure}

The code structure of \clqcd\ is displayed in \figurename~\ref{fig:code-structure}.
It is separated in two main components: the \cppnamespace{physics} package, representing high-level functionality, and the \cppnamespace{hardware} package, representing low-level functionality.
In addition, the \cppnamespace{meta} package collects what is needed to control the program execution and IO operations.

\textbf{\emph{All parts}} of the simulation code are carried out using OpenCL kernels in double precision.
The OpenCL language is based on \texttt{C99}.
In particular, concrete implementations of basic LQCD functionality like matrix-matrix multiplication, but also more complex operations like the \dslash\ or the (R)HMC force calculation, are found in the kernel files.
Their compilation and execution is handled within the \cppnamespace{hardware} package.
The kernels are in a certain way detached from the host part as the latter can continue independently of the status of the kernel execution.
This nicely shows the separation into the administrative part (host) and the performance-critical calculations (kernels). 

The \cppnamespace{physics} package provides representations of the physical objects like gauge fields or fermion fields.
In addition, the corresponding classes provide functionality to operate on the respective field type.
Moreover, algebraic operations like \texttt{saxpy} are provided.
Similarly, the various fermion matrices are provided.
This allows for the implementation of high-level functionality without knowing details of the underlying OpenCL structure.
For example, the (R)HMC or the calculation of observables are completely independent of system or kernel specifics.
In other words, the \cppnamespace{physics} package works as an interface between algorithmic logic and the actual OpenCL implementation.

In turn, the \cppnamespace{hardware} package is destined to handle the compilation and execution of the OpenCL kernels.
The \cppclass{hardware}{System} class represents the architecture available at runtime.
The latter can provide multiple computing devices (i.e. CPUs and/or GPUs), which are represented by \cppclass{hardware}{Device} objects and initialized based on runtime parameters.
Kernels are organized topic-wise within the \cppnamespace{hardware::code} namespace; for example the different fermionic fields are found in the \cppclass{hardware::code}{Fermions} classes.
These classes take over the calling logic of the kernels and provide meta informations like the number of FLOPs a specific kernels performs.
The \cppclass{hardware}{Device} class has each of the \cppnamespace{hardware::code} classes as singleton objects, i.e. they are initialized the first time they are needed.
During this process, the OpenCL kernels are compiled.

Memory management is performed by the \cppnamespace{hardware::buffers} classes, which also ensure that memory objects are treated in a \textit{Structure of arrays (SOA)} fashion on GPUs, which there is crucial for optimal memory access as opposed to \textit{Array of structures (AOS)}.

OpenCL kernels are compiled at runtime using the \texttt{OpenCL compiler} class.
In OpenCL, this is mandatory as the specific architecture is not known a priori.
On the one hand, this introduces an overhead, but on the other hand allows to pass runtime parameters (like the lattice size) as compile time parameters to the kernels, saving arguments and enabling compiler optimization for specific parameter sets.
In addition, the compiled kernel code is saved for later reuse, e.g. when resuming an HMC chain with the same parameters on the same architecture.
This reduces the initialization time.
Kernel code is common to GPUs and CPUs, device specifics are incorporated using macros.

\section{Unit Tests, Maintainability and Portability}

In general, it is desireable to be able to test every single part of code on its own and to have as little code duplication as possible.
This is at the heart of the \textit{Test Driven Development} \parencite{Beck:2002:TDD:579193} and \textit{Clean Code} \parencite{Martin:2008:CCH:1388398} concepts, which we follow during the development of \clqcd\ and which is visible in the code structure (see \figurename~\ref{fig:code-structure}). 
Unit tests are implemented utilizing the BOOST\footnote{See \myurl{http://www.boost.org}\ .} and \cmake{} unit test frameworks.

During the development of \clqcd, it was found that regression tests for the OpenCL parts are absolutely mandatory due to the runtime compilation.
The latter implies that both the architecture and the used compiler can lead to miscompilations of the kernels.
Having trustable tests at hand allows to recognize such situations quickly and simplifies error location drastically.
Most important, this can prevent the user from wasting computing time.

In particular, as LQCD functions are local in the sense that they depend only on a few nearest neighbours, one can calculate analytic results to test against.
Often, the dependence on the lattice size is easily predictable.
Varying the lattice size in the tests, or in general the parameters of the considered function, is important as errors may occur in certain parameter ranges only.

Another crucial aspects to guarantee maintainability and portability of code is to avoid dependence of the tests on specific environments.
For example, this happens when random numbers are used (e.g. for trial field configuration).
If this is the case, a test result then depends not only on the used PRNG but also on the hardware in a multi-core architecture.

\section{Performance of \dslash}

\begin{figure}[h]
  \centering
  \includegraphics[width=0.9\hsize]{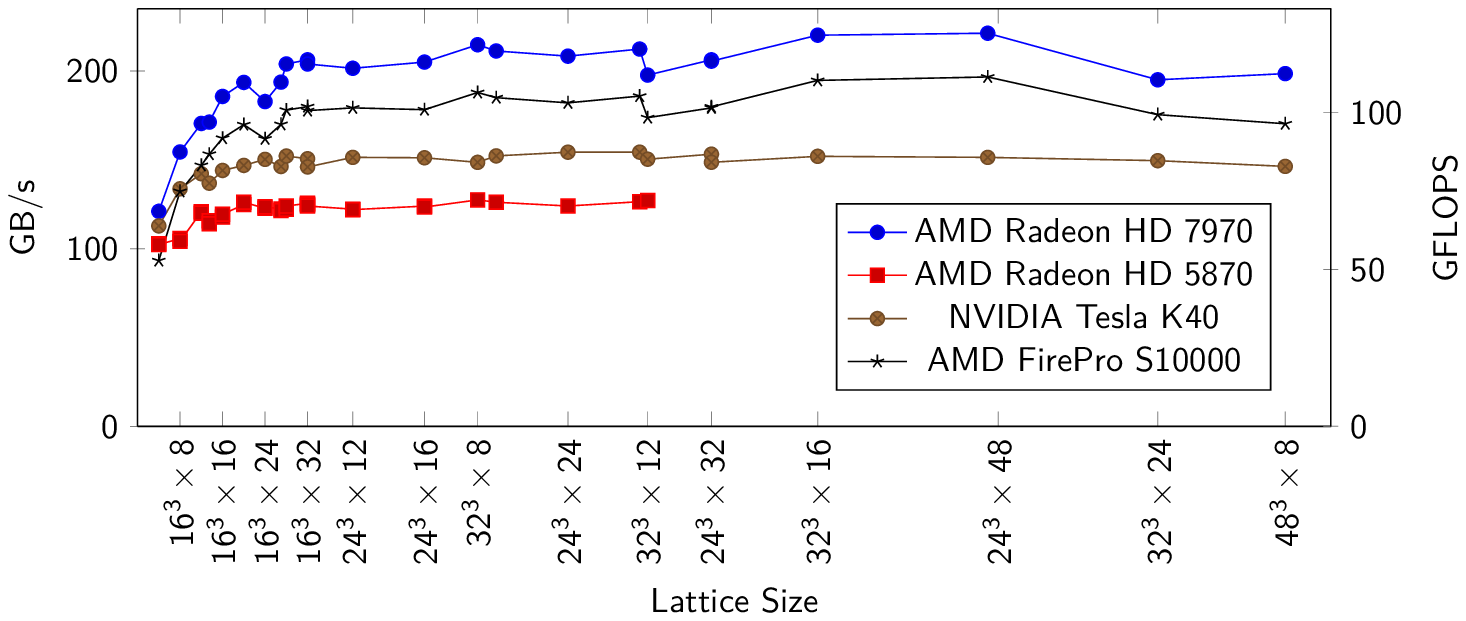}
  \caption{Performace of Wilson \dslash\ kernel for various lattice sizes on different devices in double precision.}\label{fig:dslash}
\end{figure}

Our Wilson \dslash\ implementation, which is crucial for overall performance, shows very good performance on various lattice sizes (\figurename~\ref{fig:dslash}) and outperforms performances reported in the literature (see \parencite{Bach:2012iw}). 
We are able to utilize $\sim 80\%$ of the peak memory bandwidth on the AMD Radeon HD 5870, Radeon HD 7970 and FirePro S10000.
Note that the code runs also on NVIDIA devices as shown in the figure, however, with lower performance since AMD was the primary development platform and no optimization was carried out here.

\begin{figure}[h]
  \centering
  \includegraphics[width=0.9\hsize]{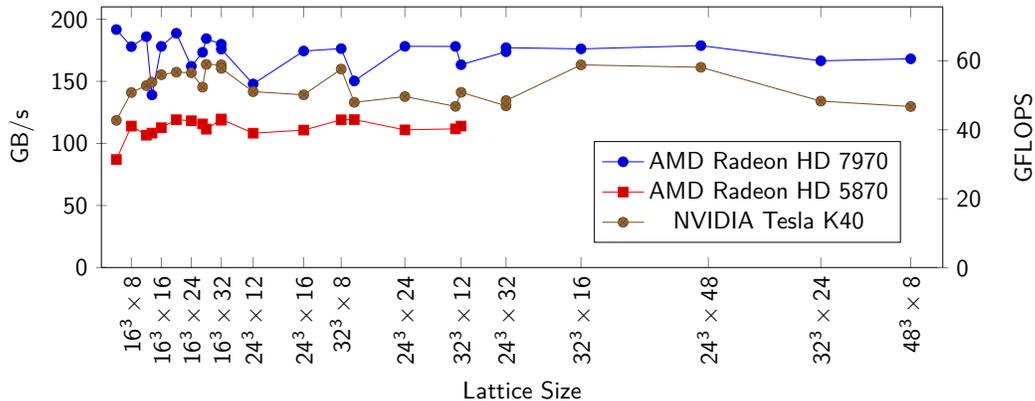}
  \caption{Performace of $D_{KS}$ kernel for various lattice sizes on different devices in double precision.}\label{fig:dks}
\end{figure}

The staggered $D_{KS}$ implementation, which plays the same role as \dslash\ regarding the overall speed of the code, shows also good performance on various lattice sizes (\figurename~\ref{fig:dks}). 
In this case we are able to utilize $\sim 70\%$ of the peak memory bandwidth on the AMD Radeon HD 5870 and AMD Radeon HD 7970.
Due to its recent development, the implementation of the staggered code can be further optimized.
So far no other benchmark for a possible comparison is present in the literature.
Again, the code runs also on NVIDIA devices as shown in the figure. The performance is though also here lower for the same reasons explained above regarding the Wilson \dslash.

\section{Algorithmic Performance}

\begin{figure}[h]
  \centering
  \includegraphics[width=.9\hsize,angle=0]{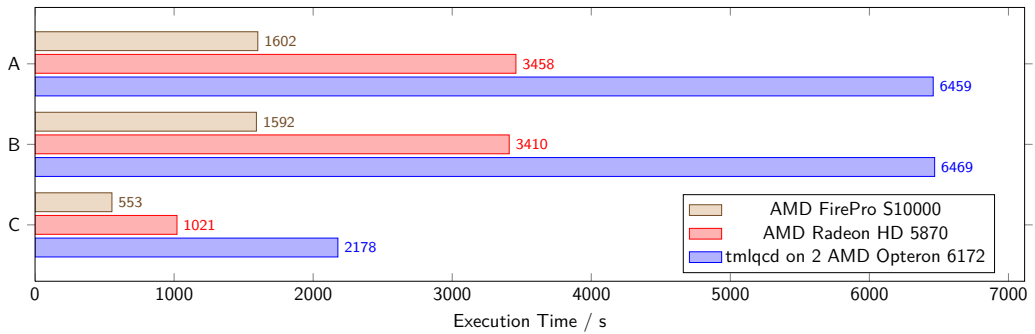}
  \caption{
     HMC performance for different Setups A, B and C (setup A having the smallest fermion mass) for $N_\tau = 8, N_\sigma = 24$. 
     The HMC is compared on different GPUs and compared to a reference code \parencite{Jansen:2009xp} running on one \loewe\ node.
  }\label{fig:hmc}
\end{figure}

The full HMC application also performs very well compared to a reference CPU-based code \tmlqcd\ \parencite{Jansen:2009xp} (see \figurename~\ref{fig:hmc}).
The \tmlqcd\ performance was taken on one \loewe\ node.
Compared to \tmlqcd, the older AMD Radeon HD 5870 is twice as fast.
The newer AMD FirePro S10000 again doubles this performance.
This essentially means that we gain a factor of 4 in speed, comparing a single GPU to a whole \loewe\ node.
In addition, it is interesting to look at the price-per-flop, which is much lower for the GPUs used then for the used CPUs. 

As on-board memory is the biggest limiting factor on GPUs, using multiple GPUs is of great interest \parencite{Babich:2011np}.
In \clqcd\ it is possible to split the lattice in time direction \parencite{Bach:2014lpa}.

\section{Conclusions and Perspectives}

We presented the OpenCL-based LQCD application {\clqcd}.
 It has been successfully applied in finite temperature studies on \loewe\ and \sanam\ supercomputers (see \tablename~\ref{tab:clusters}), providing a well-suited basis for future applications. 
\clqcd\ is available at 
\begin{center}
\begin{tikzpicture}\node[draw=black,shade,
			  top color=cyan!5!white,
			  bottom color=blue!5!white,
			  rounded corners=6pt,
			  inner sep=1ex,
			  drop shadow={
			  top color=gray,
			  bottom color=white,
			  }] {\hspace{1ex}\clqcdAdress\hspace{1ex}};
\end{tikzpicture}
\end{center}

In $N_f=2$ Lattice QCD studies we explore the phase diagram of QCD, in particular aiming at the chiral limit, where the order of the chiral transition is not resolved yet.
Results obtained here can be used to constrain the physical phase diagram of QCD.
The chiral limit is investigated in two independent approaches.
On the one hand, in studies employing Twisted Mass Wilson fermions \parencite{Philipsen:2008gq, Ilgenfritz:2009ns, Burger:2011zc}, we aim directly at the chiral limit at zero chemical potential.
On the other hand, one can approach this issue by studying the phase structure of QCD at purely imaginary values of the chemical potential $\mu$, which we do with Wilson and staggered fermions \parencite{PhysRevD.89.094504, Bonati:2013tqa}.

Additional features will be added to \clqcd\ according to the needs of the physical studies.
In the near future, these will cover the extension of Wilson fermions to $N_f = 2+1$ flavours and the implementation of the clover discretization.
Adding to that, optimizations of performances of staggered fermions and the inclusion of improved staggered actions are planned.

\acknowledgments

O. P., C. P. and A.S. are supported by the Helmholtz International Center for FAIR within the LOEWE program of the State of Hesse.
C.P. is supported by the GSI Helmholtzzentrum f\"{u}r Schwerionenforschung.
A.S. acknowledges travel support by the Helmholtz Graduate School HIRe for FAIR.

\printbibliography

\end{document}